\documentclass[12pt]{article}
\begin{document}

\title{Manipulating decay rates by entanglement and the \\ Zeno
effect.}

\author{A. F. R. de Toledo Piza$^1$ and M. C. Nemes$^2$ \\ \\
$^1$Instituto de F\'{\i}sica, Universidade de S\~ao Paulo, CP 66318,\\
05315-970 S\~ao Paulo, S.P., Brazil \\ $^2$Departamento de
F\'{\i}sica, ICEX, Universidade Federal de Minas Gerais, \\Belo
Horizonte, M.G., Brazil}

\maketitle

\begin{abstract} 

We analyse a class of quantum dynamical processes which may lead to
the hindering of the decay of a non-stationary state through
appropriate entanglement with an additional two-level system. In this
case the process can be considered as a module whose iteration is
related to dynamical implementations of the so called quantum Zeno
effect.

\end{abstract}

The dynamical evolution of (normalized) quantum-mechanical states has
the well known and peculiar feature that the evolved state becomes a
state different from its initial description only gradually, in a way
that is measured by the norm of the component of the evolved state
along the vector that describes the initial condition. The short time
behavior of this gradual process, when it is allowed to run unimpeded,
shows some general features which are particularly relevant for the
understanding of the mechanisms through which it can be modified by
interactions of the ``main'' system, in which the considered state
occurs, with other quantum systems, even in the least invasive cases
often refered to as ``non-demolishing'' interactions. More
specifically, the evolution $|\psi(t)\rangle$ of a non-stationary
state initial state $|\psi(0)\rangle$ in a quantum system whose energy
spectrum is bounded below and which has a finite mean energy is such
that the correlation $C(t)\equiv|\langle\psi(t)|\Psi(0)\rangle|^2$ has
a zero slope at $t=0$ \cite{chiu}. As shown below, this slope involves
in an essential way the coherence of the parallel and perpendicular
(or ``undecayed'' and ``decayed'') components into which
$|\psi(t)\rangle$ can be resolved. A straightforward way to reduce it,
once it develops, is therefore to quench the coherence required for
its existence.  This can be achieved by allowing for entanglement with
an external degree of freedom which has suitably arranged
properties. On the other hand, if the non stationary state has an
exponential law of decay (which implies that the mean energy is not
finite and also, strictly speaking, that the energy spectrum is not
bounded below), the decay rate is {\it not} determined by the
coherence of undecayed and decayed components, and is therefore not
affected by the entanglement process.

While the main features of the quantum dynamical processes to be
considered can easily be formulated in a more general framework, we
restrict ourselves, for definiteness, to the simple but sufficient
standard model used in the context of cavity electrodynamics, which
consists of an harmonic oscillator coupled in the rotating wave
approximation to an ``environment'' represnted by a collection of
harmonic oscillators, which corresponds to the Hamiltonian

\[
H=\omega_0a^\dagger a+\sum_k\omega_kb^\dagger_kb_k+\sum_kg_k(a^\dagger
b_k+b^\dagger_ka).
\]
 
\noindent The coupling constants $g_k$ in this context are
sufficiently small to guarantee the lower-boundedness of the spectrum
of $H$. This model Hamiltonian can be specialized to two extreme
limits of interest, namely i) the ``environment'' consists of a single
oscillator also of frequency $\omega$ and ii) the frequencies
$\omega_k$ of the ``environment'' oscillators densely cover an
extended range and are uniformly coupled to the first oscillator, so
that its initial excitation decays exponentially. Case i) can be
understood as representing a doublet of modes in a pair of identical,
ideal (very high Q-value) cavities, an arrangement in fact utilized in
ref. \cite{raimond} for different purposes. Case ii), on the other
hand, corresponds to a familiar model for a relevant mode of a single
damped cavity. In addition, use will be made os the constant of motion
$N\equiv a^\dagger a+b^\dagger b$ to restrict the discussion to the
$N=1$ (``single photon'') sector of the Hamiltonian, spanned by the
vectors $\{|10_k\rangle,\;|01_k\rangle$, in self explanatory notation.

Consider then the time evolution of the initial non-stationary state
$|10_k\rangle$, which consists of a single $\omega_0$ photon, with the
environment oscillators in their ground states. In general, this will
of course depend crucially on the distribution of environment
frequencies $\omega_k$ and coupling strengths $g_k$. Prior to any
specialization, however, it is useful to note that the time evolved
state can be generally cast in the form

\[
|t\rangle=e^{-iHt}|t=0\rangle=A(t)|10_k\rangle+|O(t)\rangle.
\]

\noindent Note that $|O(t)\rangle$ is not normalized, having been
defined as the component orthogonal to $|10_k\rangle$ of the
corresponding time-evolved state, i.e

\[
|O(t)\rangle=(1-|10_k\rangle\langle 10_k|)e^{-iHt}|10_k\rangle,
\]

\noindent while $A(t)\equiv\langle 10_k|e^{-iHt}|10_k\rangle$. The
probability of finding the $\omega_0$ photon is now given, at a
somewhat latter time $t+\tau$, can be cast in the form

\begin{eqnarray}
\label{Pcav}
P(t+\tau)&=&|A(t+\tau)|^2=\nonumber \\ &=&|A(t)|^2|A(\tau)
|^2+|B_t(\tau)|^2+2{\rm Re}\,\left[A(t)A(\tau)B_t^*(\tau)\right]
\end{eqnarray}

\noindent with the definition

\begin{equation}
\label{Bttau}
B_t(\tau)\equiv\langle 10_k|e^{-iH\tau}|O(t)\rangle=A(t+\tau)-
A(t)A(\tau).
\end{equation}

In order to calculate the decay rate $-dP(t+\tau)/d\tau$ at $\tau=0$
one must now specify the characteristics of the environment and of its
coupling to the $\omega_0$ oscillator in some detail. In fact, for
small $t$ one has formally 

\begin{equation}
\label{ashortt}
A(t)\simeq 1-it\langle 10_k|H|10_k\rangle 
\end{equation}

\noindent which is certainly meaningful in the case in which the
environment is modeled by a finite number of oscillators (a weaker
condition being that the initial state $|10_k\rangle$ has a finite
mean energy). This behavior of $A(t)$ is sufficient to guarantee that
the only contribution to the decay rate comes from the interference
term in eq. (\ref{Pcav}), and is given by

\begin{equation}
\label{decrate}
\left.-\frac{dP(t+\tau)}{d\tau}\right|_{\tau=0}=-2{\rm Re}\left.
A(t)\frac{dB_t^*(\tau)}{d\tau}\right|_{\tau=0},
\end{equation}

\noindent in general a finite, nonvanishung quantity. This decay rate
can be made to vanish by entangling the state $|\psi(t)\rangle$ with
orthogonal states $|\alpha_I\rangle$ and $|\alpha_O\rangle$ of an
external two-level system (e.g. a two-level atom), previously in some
state $|\alpha_p\rangle$, as

\begin{equation}
\label{dec}
|t\rangle\otimes|\alpha_p\rangle\longrightarrow A(t)|10_k\rangle
\otimes|\alpha_I\rangle+|O(t)\rangle\otimes|\alpha_O\rangle.
\end{equation}

\noindent The states $|\alpha_I\rangle$ and $|\alpha_O\rangle$ are
then assumed to evolve independently and unitarily with an operator
$U_\alpha(t)$ (so that their orthogonality is preserved in time). No
mesurement is assumed to be performed on these states. The interaction
time for the entanglerment process has been assumed to be very short
in the sense that the effects of $H$ can be ignored. The interaction
leading to the entanglement can be thought of as a non-invasive,
impulsive interaction of the von Neumann type, so that it merely
correlates diferent components of the state representing the cavities
to different atomic states, without affecting the former
otherwise. Appart from questions relating to the time duration of the
interaction process, effects precisely of this sort have been devised
in atom-cavity interactions using interactions rendered dispersive by
detuning \cite{brune} or, more recently, the so called SP-QND (for
single photon quantum non-demolition) {\it resonant} procedure
\cite{nogues}. 

We next evaluate the decay rate for the photon in the state
(\ref{dec}), {\it subsequent} to the interaction with the atom,
and assuming that the atom itself remains unobserved. To this effect
we again let the system evolve freely for sometime $\tau$, which now
changes the state into

\[
A(t)e^{-iH\tau}|10_k\rangle\otimes U_\alpha(\tau)|\alpha_I\rangle+
e^{-iH\tau}|O(t)\rangle\otimes U_\alpha(\tau)|\alpha_O\rangle.
\]

\noindent The procedure to obtain $P(t+\tau)$ consists now in
projecting this state successively onto $|10_k\rangle\otimes|\alpha_I
\rangle$ and $|10_k\rangle \otimes|\alpha_O\rangle$, and summing the
corresponding probalilities {\it incoherently}. The result

\begin{equation}
\label{Pent}
P_{\rm ent}(t+\tau)=P(t)P(\tau)+|B_t(\tau)|^2
\end{equation}

\noindent reproduces that obtained in equation (\ref{Pcav}) {\it
without the cross-term contribution}. It follows that, as a
consequence of the entanglement process with the two-level system, the
decay rate is reset to zero at the time of the interaction.

These results give rise to very simple closed expressions in case
i), of two identical, ideal coupled cavities. In this case, in fact,
eqs. (\ref{Pcav}) and (\ref{Pent}) become

\[
P(t+\tau)=\cos^2g(t+\tau)
\]

\noindent and

\[
P_{\rm ent}(t+\tau)=P(t+\tau)+\frac{1}{2}\sin2gt\;\sin2g\tau.
\]

\noindent Therefore one gets, through the decoherence process, an
excess probability proportional to $\sin2g\tau$. In the particular
case when $gt=\pi/4$ the result for $P_{\rm ent}$ reduces to $P_{\rm
ent}(\pi/4g+\tau)=\frac{1}{2}$. It is then independent of $\tau$,
indicating that in this case the excess probability just makes up for
the further decay of $P(\pi/4g+\tau)$. An alternate way of expressing
this result is to calculate the decay rate associated with $P_{\rm
ent}$ as a function of $\tau$, which is given by

\[
-\frac{d}{d\tau}P_{\rm ent}(t+\tau)=g\cos2gt\sin2g\tau,
\hspace{1cm}\tau>0
\]

\noindent which vanishes identically for $gt=\pi/4$ and in
fact becomes {\it negative} for somewhat larger values of $t$.

The above results can be interpreted in a rather straightforward way
on the basis of the realization that imediately after the entanglement
process the state of the coupled cavities is described as an
incoherent superposition of the localized states $|10\rangle$ and
$|01\rangle$, with weights given respectively by $\cos^2gt$ and
$\sin^2gt$. This state corresponds to the density operator

\[
\rho(t)=|10\rangle\cos^2gt\langle 10|+|01\rangle\sin^2gt\langle 01|.
\]

\noindent Each of the components of this density operator evolves
through the dynamics prescribed by the Hamiltonian $H$. While the
$|10\rangle$ component decays to the $|01\rangle$ component, this
latter component in turn back-feeds the $|10\rangle$ component. For
$gt=\pi/4$ the two components have the same initial weight,
and these two processes just compensate each other. For somewhat
larger values of $t$, the $|01\rangle$ component dominates, and
back-feeding outweights the decay of the $|10\rangle$ component.

Case ii) gives rise to an entirely different picture. The amplitude
$A(t)$ can be written using the expansion of the initial state in the
$N=1$ eigenstates $|\Omega_n\rangle$ of $H$, $|10_k\rangle=\sum_na_n|
\Omega_n\rangle$ as

\[
A(t)=\sum_n|a_n|^2e^{-i\Omega_nt}.
\]

\noindent where the $\Omega_n$ are the eigenfrequencies. An
exponential decay law $A(t)=e^{-\Gamma t/2}$ results when the
$|a_n|^2$ can be well described by a distribution of the form

\begin{equation}
\label{BW}
|a_n|^2\longrightarrow\frac{1}{2\pi}\;\frac{\Gamma d\Omega}
{(\Omega-\Omega_0)^2+\Gamma^2/4},\hspace{.5cm}\Omega_0\approx\omega_0
\end{equation}

\noindent which is realistic when the environment oscillators are
uniformly distributed in frequency and uniformly coupled to the
microwave cavity mode with $g_k=g\ll \omega_0$. In this case
$\Gamma=2\pi g^2\,dn/d \Omega$. It should be noted that these
assumptions ammount to the choice of an initial state of the {\it
coupled} cavity plus environment system which does not have a finite
mean energy. In this case $A(t+\tau)=A(t)A(\tau)$ and, by virtue of
eq. (\ref{Bttau}), $B_t(\tau)=0$. This implies

\[
-\left.\frac{dP(t+\tau)}{d\tau}\right|_{\tau=0}=\Gamma e^{-\Gamma t}
\]

\noindent a result which is unaffected by entanglement with the
two-level system. It may be noted that, strictly speaking, the
distribution (\ref{BW}) in not consistent with an energy spectrum
which is bounded below, as is the case of the relevant realizations of
$H$. This means, also strictly speaking, that an exponential decay law
is incompatible with this condition on the spectrum. For practical
purposes, howeves, deviatiations can be unobservably small, in which
case one still has $B_t(\tau)\approx 0$.

It is easy to appreciate at this point that the hindrance mechanism of
decay rates discussed above hinges crucially, on the one hand, on the
distribution of the decaying state on the energy spectrum of the host
system and, on the other hand, on the particular nature of the
involved entanglement process. Concerning the spectral distribution of
the decaying state, the conditions of finite mean energy and lower
bound leads to the vanishing of the decay rate at $t=0$. If, in
addition, the decaying state has a finite mean squared energy,
$\langle H^2\rangle<\infty$, one has $P(t)\propto t^2$ for short
times, as e.g. in case i) above. The latter condition actually
characterizes the states belonging to the natural domain of the
considered Hamiltonian.

The entanglement process which hinders decay rates which are initially
vanishing must be such that it destroys the coherence of the
``undecayed'' and ``decayed'' components of the initial state, since
this coherence is responsible for the growth of the initially
vanishing decay rate, as shown by eq. (\ref{decrate}). It should be
stressed that one is here dealing with the state of the {\it joint}
system consisting e.g. of the relevant ``cavity mode'' {\it and} its
environment. This effect is therefore inaccessible to a master
equation description of the dynamics of the first of these subsystems
alone, since in this type of description the correlations between the
two subsystems are absent by construction. This feature is of course
related to the fact that the latter type of treatment leads always to
exponential decay laws.

If the interaction time $\Delta t$ for the entanglement process cannot
be ignored, a relation similar to eq. (\ref{dec}) still holds, with
the coefficient $A(t)$ repalced by some appropriate $A'(t+\Delta t)$
and with the component $|O(t)\rangle$ replaced accordingly with $|O'
(t+\Delta t)\rangle$, following which the decay rate again vanihes. No
qualitative differences result therefore from this time delay,
provided it is sufficiently small. In fact it is shown in
ref. \cite{ze} that, using the SP-QND scheme in connection with a
situation akin to case ii) above, one has $|A'(t+\Delta t)|^2>|A(t+
\Delta t)|^2$, meaning that the probability of having one photon in
the cavity after the resonant interaction exceeds that which would
result should the system be allowed to evolve by itself. This
indicates that, in a situation of the type i), that one has a {\it
temporary} inhibition of the decay process associated with the
interaction responsible for the entanglement, in addition to the
subsequent effect of the latter.

The modification of a decay rate by entanglement with an external
subsystem can be seen as a ``dynamical module'' for the quantum Zeno
effect, as described by Pascazio and Namiki \cite{pasc}. Refering
specifically to case i) discussed above, with some additional
algebraic work it is possible to calculate the probability
$P^{(N)}(N\delta t)$ of finding the coupled pair of oscillators in
their initial state $|10\rangle$ after N iterations of the impulsive
module. The iterations are separated by equal time intervals $\delta
t$ during which the oscillators evolve with the Hamiltonian $H$, and
each one of the N entanglement processes involves a different,
independent two-level system. The result of this calculation is

\begin{equation}
\label{inclN}
P^{(N)}(N\delta t)=\frac{1}{2}\left(1+\cos^N2g\,\delta t\right).
\end{equation}

\noindent It should be noted that this probability {\it includes} the
feeding of the $|10\rangle$ components of the manifold entagled state
by the previously ``decayed'' $|01\rangle$ components. In the limit
$N\rightarrow\infty$ with $\delta t=t/N\rightarrow 0$, $t$ being some
finite overall time interval, one gets $P^{(N)}(t)\rightarrow 1$,
which corresponds to the freezing of the decay process.

The result (\ref{inclN}) differs from the probability obtained by the
straightforward application of the simple product rule, which gives

\begin{equation}
\label{redN}
Q^{(N)}(N\delta t)\equiv\left[P^{(1)}(\delta t)\right]^N=\cos^{2N}
g\,\delta t.
\end{equation}

\noindent This alternate calculation corresponds to observing the
final state of the two-level system after each one of the entanglement
interactions and filtering out the decayed outcomes, which is
equivalent to successive applications of the reduction postulate.
While the above described limit of the probability $Q^{(N)}(N\delta
t)$ is also unity, it is interesting to observe that (\ref{inclN}) and
(\ref{redN}) lead to different results when the limit $N\rightarrow
\infty$ is taken with a fixed value of the repetition time interval
$\delta t<\pi/4g$. In this case $Q^{(N)}(N\delta t)$ decreases to zero
while, as a result of the back-feeding of the $|10\rangle$ state by
the previously decayed $|01\rangle$ components, $P^{(N)}(N\delta t)$
decreases monotonically to the value $1/2$. It is also worth noting
that these two procedures lead to different results for any $N\geq 2$.

\end{document}